\documentclass{aastex}
\usepackage{spr-astr-addons}
\usepackage{url}

\def\mbh{$M_{\rm BH}$\/}
\def\nh{$n_{\mathrm{H}}$\/}

\def\rfe{$R_{\rm FeII}$}

\def\feiiq{\rm Fe{\sc ii}$\lambda$4570\/}

\def\ltsima{$\; \buildrel < \over \sim \;$}
\def\ltsim{\lower.5ex\hbox{\ltsima}}  
\def\simlt{\lower.5ex\hbox{\ltsima}}  
\def\gtsima{$\; \buildrel > \over \sim \;$}

\def\gtsim{\lower.5ex\hbox{\gtsima}} 
\def\simgt{\lower.5ex\hbox{\gtsima}}

\def\lya{{ Ly}$\alpha$}
\def\civ{{\sc{Civ}}$\lambda$1549\/}

\def\cm3{cm$^{-3}$\/}
\def\hb{{\sc{H}}$\beta$\/}

\def\mgii{{Mg\sc{ii}}$\lambda$2800\/}

\def\ciii{{\sc{Ciii]}}$\lambda$1909\/}

\def\o4363{{\sc{[Oiii]}}$\lambda$4363\/}

\def\oiiiuv{{\sc{Oiii]}}$\lambda$1663\/}
\def\siiii{Si{\sc iii]}$\lambda$1892\/}
\def\aliii{Al{\sc iii}$\lambda$1860\/}
\def\heiiuv{He{\sc{ii}}$\lambda$1640}
\def\nv{{N\sc{v}}$\lambda$1240}

\def\feii{{Fe\sc{ii}}\/}
\def\siii{{Si\sc{ii}}$\lambda$1814\/}
\def\feiii{{Fe\sc{iii}}\/}

\def\fe{{\sc{Fe}}\/}

\def\fe76087{{\sc [Fe vii]}$\lambda$6087\/}

\def\kms{km~s$^{-1}$}

\def\heii{{{\sc H}e{\sc ii}}$\lambda$4686\/}

\def\siiv{Si{\sc iv}$\lambda$1397\/}
\def\oiv{O{\sc iv]}$\lambda$1402\/}

\def\niii{N{\sc iii}]$\lambda$1750\/}

\def\oi{O{\sc i}$\lambda$1304\/}

\RequirePackage{color}

\begin{document}

\title{UV spectral diagnostics for  low redshift quasars:
estimating  physical conditions and radius of the Broad Line Region}
\shorttitle{Short article title}
\shortauthors{Autors et al.}

\author{P. Marziani}
\affil{INAF, Osservatorio Astronomico di Padova, Padova, Italia}
\and 
\author{J. W. Sulentic}
\affil{Instituto de Astrof{\'\i}sica de Andaluc{\'\i}a (CSIC),    Granada, Spain}
\and
\author{C. A. Negrete}
\affil{INAOE,   Tonantzintla, Puebla,    Mexico}
\and
\author{D. Dultzin}
\affil{Instituto de Astronom{\'\i}a, UNAM, 
Mexico, D.F.,   Mexico}
\and
\author{A. Del Olmo}
\affil{Instituto de Astrof{\'\i}sica de Andaluc{\'\i}a (CSIC),    Granada, Spain}
\and
\author{M. A. Mart\'{\i}nez Carballo}
\affil{Instituto de Astrof{\'\i}sica de Andaluc{\'\i}a (CSIC),    Granada, Spain}
\and
\author{T. Zwitter}
\affil{University of Ljubljana, Faculty of 
Mathematics and Physics, Ljubljana, Slovenia}
\and
\author{R. Bachev}
\affil{Bulgarian Academy of Science, Sofia, Bulgaria}



\begin{abstract}
The UV spectral range (1100 -- 3000 \AA)  contains the strongest resonance lines observed in active galactic nuclei (AGN).  
Analysis of UV line intensity ratios and profile shapes  in quasar spectra provide diagnostics of physical and dynamical 
conditions in the broad line emitting region.  This paper discusses properties of UV lines in type-1 AGN spectra, 
{ and how they lead an estimate of ionizing photon flux,  chemical abundances, radius of the broad line emitting 
region and   central black hole mass. These estimates are meaningfully contextualised  through the   
4D ``eigenvector-1" (4DE1) formalism.}
\end{abstract}
\keywords{galaxies: active; quasars: emission lines; quasars: general; techniques: spectroscopic; astronomical databases: surveys}


\section{Introduction}

It is now roughly 35 years since the International Ultraviolet Explorer (IUE) opened the possibility to study  UV spectra of  
Seyfert nuclei and broad line radio galaxies thus revealing  strong similarities between low- and high-redshift quasars (where 
the UV lines are redshifted into the optical domain). IUE data helped  to establish that active galactic nuclei (AGN) involved 
sources spanning an enormous range of luminosity but driven by the same underlying process involving accretion onto 
a massive compact object, most likely a black hole (see e.g., Chapter 1 of \citealt{donofrioetal12}). The advent of HST in 1991 made 
possible observations of even fainter AGN and with unprecedented resolution (comparable to the best ground based optical 
spectra). HST archival observations continue to provide a database from which much AGN research is being carried out.  
Important information about density, ionization conditions, and dynamics in the (especially  high-ionization) broad line 
emitting region (HIL--BLR) of AGN can be inferred from UV spectroscopic observations (\S \ref{diag} and \S \ref{results}). 
{   It is useful to point out three factors affecting our ability to achieve these goals. 
\begin{itemize}
\item  The line emitting regions in type-1 AGN are spatially  unresolved in even the nearest AGN.  This is an important reason 
why  modeling the structure of the inner  regions of quasars remains an open issue (see \citealt{netzer13}  
and \citealt{gaskell09}).  Much effort has gone  into measuring the size of the broad line emitting regions (BLR)  from 
time lags in the response of emission lines to continuum change \citep{petersonetal04}.  Work has also focussed on constraining 
the BLR  structure using  velocity-resolved emission line profiles \citep[][and references therein]{grieretal13}.

\item    AGN spectral properties show considerable diversity. 
The point is that  type-1 AGN spectra are not self-similar nor do they scatter with low  dispersion around an average spectrum 
\citep{sulenticetal02,bachevetal04}. If they did then we would not need { the  contextualization described in \S \ref{4de1}}, just as we would not need an H-R 
diagram if all stars were of the same spectral type. Type-1 AGN  actually show large multi-wavelength diversity ranging from 
Narrow Line Seyfert 1 (NLSy1) to Broad Line Radio Galaxies (BLRG).  These two extremes involve sources whose emitting 
regions are empirically and physically very different.  Attempts to explain  them with a single model will yield misleading 
results \citep{sulenticetal00b,sulenticetal02}.

\item   Finally we point out that the availability of UV spectra is limited and that this is an impediment to further progress in quasar  studies involving comparisons of low- and high-$z$\  AGN  (\S\ref{mast}). { This concerns  single epochs observations suitable for statistical studies as well as monitoring for reverberation mapping. }
\end{itemize}

\section{The UV Diagnostics}
\label{diag}

\subsection{Prominent UV emission lines}

An important property of AGN spectra involves the existence of both  high and low ionization lines (hereafter HILs and LILs).
By low and high ionization we mean, respectively,  lines emitted by ionic species with ionization potential $\la$20 eV (hydrogen, singly 
ionized ionic species of magnesium, carbon, iron, calcium) and $\ga$ 40 eV (triply ionized carbon, helium, four times ionized 
nitrogen). An exhaustive list of lines  is provided by \cite{vandenberketal01}. If we confine our studies to  the range 1100 -- 2000 \AA, 
we observe: (1) the strongest HILs associated with resonance transitions, (2) usually weak  LIL emission due to \oi, \siii\ features and  (3) 
several inter-combination lines from transitions leading to the ground state, most notably, \ciii, \siiii, and the narrow lines of \niii\ and \oiiiuv.   
Once and twice ionized iron emission is not strong between 1200 and 2000 \AA\ although \feiii\ emission can be significant in 
the 1900 \AA\ blend { whose main components are \aliii, \siiii, and \ciii} \citep{vestergaardwilkes01}.  The most notable features that 
are unblended with stronger lines include the UV multiplet 191 \feii\ at 1780 \AA\ and an \feiii\ blend at 2100 \AA. It is customary to assume 
that \feii\ and, to some extent, \feiii\ emission maintain the same intensity ratios in all sources. Under this assumption  intensity measures 
of  \feii\ and \feiii\ at 1780 and 2100 \AA\ define  the entire UV emission spectrum due to these ions. This approach (i.e., using a ``template'' 
to represent blended emission features spread over a wide wavelength range) is justified within the limit of resolution and S/N presently 
obtainable  for most optical and UV data although the real situation is known to be more complex. 

\subsection{Line Ratios}

Line intensity ratios are sensitive to: (1) density,  if one of the lines has a well-defined critical density, (2) ionisation level, with   { lines coming from transitions occurring in different ionic species of the same element providing the most  robust diagnostics}, and (3) chemical composition, if ratios involve a metal line and \heiiuv\ { or \civ}. 
Intensity ratios of broad inter-combination and permitted lines, most notably   C{\sc iii}$\lambda$1909/Si{\sc iii}]$\lambda$1892\ and 
Al{\sc iii}$\lambda$1860/ Si{\sc iii}]$\lambda$1892\  are useful diagnostics for a range of density that depends on their transition probabilities. 
The  C{\sc iii}$\lambda$1909/ Si{\sc iii}]$\lambda$1892 and  Al{\sc iii}\-$\lambda$1860/Si{\sc iii}]\-$ \lambda$1892\ ratios are suitable as   
diagnostics for   $n_\mathrm{e} < 10^{11}$ cm$^{-3}$\  and $10^{11} - 10^{13} $ cm$^{-3}$ ranges, respectively. The latter range 
corresponds to the densest, low ionization emitting regions likely associated with production of Fe{\sc ii} \citep{sigutpradhan03,bruhweilerverner08}.  
The  ratios  Si{\sc ii}]\-$\lambda$1814/Si{\sc iii}]$\lambda$1892 and Si{\sc iv}$\lambda$1397/ Si{\sc iii}]$\lambda$1892  are sensitive 
to the ionization level and have the considerable advantage of being independent of chemical abundance. Ratios 
C{\sc iv} $\lambda$1549/ Al{\sc iii} $\lambda$1860 and C{\sc iv} $\lambda$1549/ Si{\sc iii} $\lambda$1892 provide information about 
ionization that is however metallicity dependent.  The  ratios involving \nv\ (\nv/\heiiuv\ or  \nv/\civ) are strongly sensitive to 
metallicity if nitrogen is considered  a secondary element \citep{ferlandetal96}. The ratio (\siiv + \oiv) / \civ\   has also been used as a  
metallicity indicator \citep{nagaoetal06b}.   

\subsection{Photoionization calculations}

The emission line ratios reported above provide information about the product of ionization parameter $U$\ and hydrogen number density $n_\mathrm{H}$\ if 
interpreted through  a multidimensional grid of  photoionization simulations.  Quantitative constraints obtained from diagnostic ratios require  a-priori knowledge of the principal ionization mechanism and some 
basic assumptions about the structure of the emitting region. Photoionization is considered the dominant mechanism for 
HIL and intermediate ionization lines \citep{davidsonnetzer79}. Physical conditions of the photoionized gas can be described by: 
1)    density $n_\mathrm{H}$\ or   electron density, 2)   hydrogen column density $N_\mathrm{H}$\, 3)   
metallicity $Z$, 4)   shape of the ionizing continuum and 5)   ionization parameter $U$.  $U$ represents the dimensionless 
ratio of the number of ionizing  photons and the total hydrogen density. Both $U$\ and $n_\mathrm{H}$\ are related through 
the equation
\begin{equation}
\label{eq:u}
U = \frac {\int_{\nu_0}^{+\infty}  \frac{L_\nu} {h\nu} d\nu} {4\pi n_\mathrm{H} c r^2}
\end{equation}
where $L_{\nu}$\ is the specific luminosity per unit frequency, $h$\ is  the Planck constant, $\nu_{0}$\ the Rydberg frequency, 
$c$ the speed of light, and $r$\ can be interpreted  as the distance between the central source of ionizing radiation and the line 
emitting region.  Simulations with {\sc cloudy} \citep{ferlandetal13} were 
carried out over the density range $7.00 \leq \log n_\mathrm{H} \leq 14.00$, and $-4.50 \leq \log U \leq 00.00$ (in intervals of  0.25).  
Each simulation was computed for a fixed ionization parameter and density, assuming plane parallel geometry (i.e., a slab of gas 
illuminated by the ionizing continuum).  The $n - U$\ array was  computed  several times for solar and supersolar chemical compositions, 
two different continua and  column density values $N_\mathrm{H} = 10^{23}$\ (assumed as a standard value), 
plus $N_\mathrm{H} = 10^{22}$, $10^{24}$\ cm$^{-2}$.

Fig. \ref{fig:contours} shows the behaviour of \siiii / \civ\  (left panel) and \siiii / \aliii\ (right panel) in the $U$ vs. $n_\mathrm{H}$ plane assuming solar metallicity, $N_\mathrm{H} = 10^{23}$ cm$^{-2}$, and the standard {\sc cloudy} AGN continuum. As expected 
the  \siiii/\civ\   ratio depends on $U$ in a way that is almost independent of density up to the  \siiii\ critical density. The \aliii/\siiii\ ratio 
increases smoothly with density with an increase that is not strongly dependent on $U$\ for $-3.5 \la U \la -1$.

\begin{figure*}[t]
\includegraphics[scale=0.5]{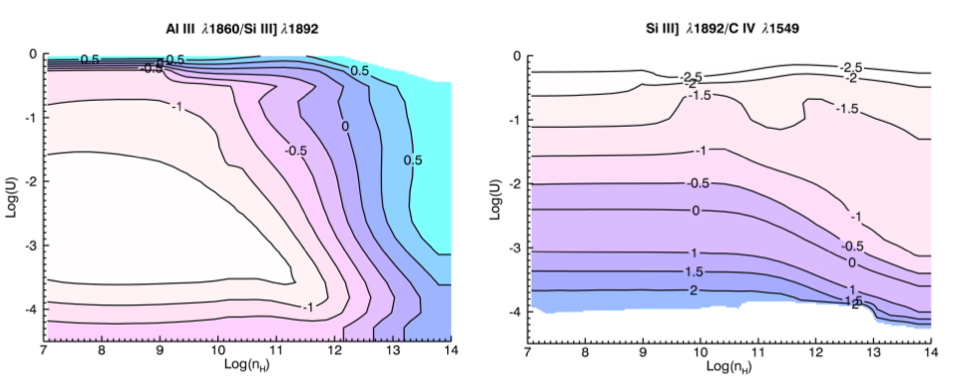}
\caption{Behavior of  $\log$ \siiii/\civ\ and $\log$  \aliii/\siiii\ in the 2D parameter space defined by ionization parameter $U$\ and hydrogen 
density $n_\mathrm{H}$, from a set of 551 simulations  that assume solar metallicity, a standard AGN continuum, and $N_\mathrm{c} = 10^{23}$\ cm$^{-2}$. { Two arrays of the \citet{negreteetal12} simulations are shown here to illustrate two cases in which a single line ratio defines a clear trend:  \aliii/\siiii\ for density, \siiii/\civ\   for ionization parameter   (if density is less than the critical density of \siiii). }}  \label{fig:contours} 
\end{figure*}

\section{Systematic changes along the 4DE1 sequence}
\label{4de1}\label{results}

Until recently we  lacked a contextualization within which the  dispersion in optical and UV  spectroscopic properties could be  organized (e.g. a quasar equivalent of the stellar H-R diagram).  There has been recent progress  especially if we restrict our attention to type-1 AGN.   Spectroscopic measures from large AGN samples can be viewed as  defining a   parameter space whose axes are not fully orthogonal.  A principal component analysis (PCA) can be used to identify the  first orthogonal axes (which are  expressed as linear combinations of the observed parameters) that account for most of the parameter 
diversity in a sample.   The ``first eigenvector"  revealed by PCA organizes  AGN along a sequence originating in an optical
parameter plane FWHM \hb\ vs. intensity ratio  \rfe = \feiiq/\hb\   \citep{borosongreen92}. 
{  This is the  ``optical" plane of the 4D eigenvector 1 parameter space \citep{sulenticetal00b} that adds  systematic trends 
involving  \civ\ and the soft X-ray photon index \citep{sulenticetal07}.  If trends in physical conditions are taken into account  it is possible to obtain   accurate estimates of the emitting region radius  from this data (\S \ref{rm}). }

 { Systematic differences along the the eigenvector 1 sequence do not involve only  \feii\ and FWHM \hb\ \citep{zamfiretal10,popovickovacevic11,steinhardtsilverman13},  but also  the spectral energy distribution \citep{kuraszkiewiczetal09,tangetal12}, 
X-ray continuum / \civ\ properties \citep{kruzceketal11}, the ratio between optical and UV \feii\ emission \citep{sameshimaetal11}, and, most notably,  the ratios \aliii / \siiii, and \siiii/ \ciii\ \citep{baldwinetal96,willsetal99,bachevetal04}.  { Eddington ratio  is considered the main physical driver of the eigenvector 1, with orientation affecting mainly line widths \citep{marzianietal01,boroson02}. } \citet{ferlandetal09} establish a link between \rfe, column density and Eddington ratio along the eigenvector 1 sequence.

Several authors  distinguish between two main type-1 AGN populations that involve sources at the opposite ends of the eigenvector 1 sequence, and  speak of NLSy1s and broad(er) line quasars (BLQs) or  Population A and B(roader) \citep{sulenticetal00b} or Population 1 and 2  \citep{collinetal06} or disk- and wind-dominated sources \citep{richardsetal11}. We adopt the Pop. A and B subdivision involving a boundary   at FWHM(\hb) $\approx$ 4000 \kms\ \citep[][and references therein]{sulenticetal07}.

\begin{figure*}[t]
\begin{center}
\includegraphics[scale=0.5]{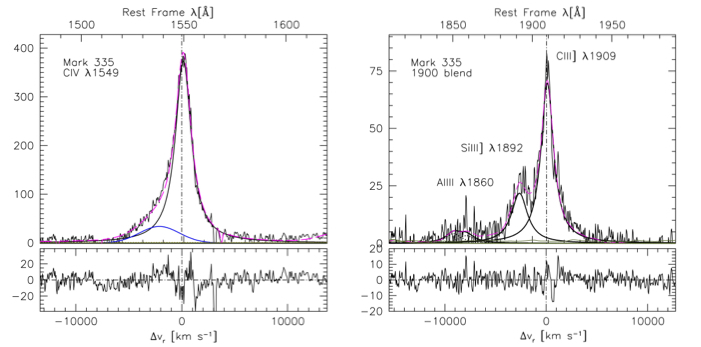}
\caption{Analysis of relevant blends  for Mark 335. Left panel: the \civ\ line profile is fit using a scaled \hb\ profile 
(black line), symmetric and almost unshifted whose width can serve as estimator of virial broadening; the residual 
on the blue side is the excess component attributed to a wind or outflow. Right panel: the thick lines represent   
the  features in the blend: doublet \aliii, and single lines \siiii, \ciii. UV \feii\ and \feiii\  (grey line close to 0) emissions are 
barely detectable in this object. \label{fig:prof}} 
\end{center}
\end{figure*}

\subsection{Dynamics of the Broad Line Regions}

In order to properly  interpret  intensity ratios  it is  necessary to understand that   HILs  and LILs arise from emitting gas not necessarily in the same physical and dynamical conditions \citep[e.g.,][]{sulenticetal07,richardsetal11}.  HST has made it possible to measure  line shifts with an accuracy of $\pm  200$ \kms\ relative to a quasar rest frame derived from optical observations.  The prototypical HIL  \civ\ and other HILs frequently show blueshifts that can reach several thousand \kms\ \citep[]{gaskell82,tytlerfan92,brothertonetal94,marzianietal96}  indicating the presence of winds or outflows.  
Dynamics  conditions change along the  4DE1 sequence \citep{bachevetal04} with the HIL blueshift becoming weaker as one proceeds to  less  \feii\ strong quasars with  broader lines. 


\subsection{Physical conditions of the BLR}

Line widths increase with increasing ionization potential  and higher ionization lines usually respond more rapidly to continuum changes. This  
indicates an ``ionization stratification''  within the BLR \citep{petersonwandel99}. The stratification is also empirically confirmed by the simultaneous 
presence of strong \ciii\ and \siiii\ representing transitions involving critical densities differing by more than one order of magnitude. 
This makes the search for a single solution in terms of density and ionization rather ambiguous.  Nonetheless, Eq. \ref{eq:u} can be used 
to estimate the ionizing photon flux, i.e., the product $n_\mathrm{H} U$, as we will discuss below. 

 \citet{negreteetal13} studied a sample of 13 type-1 AGNs with high-quality HST/FOS coverage 
in  the range 1200 -- 2000 \AA, and with $c \tau$\ measures for \hb\ from monitoring campaigns \citep{bentzetal13}. 
These authors obtained two different solutions for $n_\mathrm{H} U$\ depending on whether  \ciii\ or \aliii\ was used: in the first case 
the solution involved high ionization and low density and in the second  high density and low ionization. 
These solutions were compared with the ones derived using reverberation mapping i.e., setting $c \tau \approx r_\mathrm{BLR}$ in 
Eq. \ref{eq:u}, where the time lag $\tau_\mathrm{L}$ represents the light travel time for continuum photons to reach the broad line region. Only the high-density solution showed good  agreement with $c\tau_\mathrm{L}$. \ciii\ yields systematically lower $n_\mathrm{H} U$
which is  a result that can be easily understood in terms of stratification within the BLR or other more complex scenarios \citep{maiolinoetal10}.

\subsection{Physical conditions at the 4DE1 extremes}

At the extremes of the 4DE1 sequence  physical conditions appear better defined and it is possible to derive a typical density 
and ionization parameter for the BLR. At the high \rfe\ end of the 4DE1 sequence (extreme Population A) we find sources 
whose optical spectra are dominated by LIL emission. In the UV such sources  show large HIL blueshifts and weak \ciii. 
They are interpreted as sources radiating close to the  Eddington limit with HILs mostly reflecting high-ionization winds. Using the 
diagnostic line ratios described  in \S \ref{diag}, \citet{negreteetal12} find well defined physical conditions:  low ionization ($U \la 10^{-2}$)
and high density (10$^{12} - 10^{13}$ \cm3) with an uncertainty less than $\pm$ 0.3 dex along with significant metal enrichment. 
The weakness of \ciii\ enables all important  emission line ratios to converge toward a  high-density solution. Extreme \rfe\ sources 
show low equivalent width in {\em all} the strongest broad  emission lines i.e, C{\sc iv}$\lambda$1549 and  Balmer lines: 
extreme Eddington ratio sources may have stripped the BLR of lower column density gas, leaving only the densest part of the BLR  
\citep{negreteetal13}.   These sources show extreme iron and aluminium emission which  
might be associated with chemical enrichment \citep{juarezetal09}; selective aluminium and silicon abundance 
enhancement may be needed to explain in detail the observed emission line ratios   \citep{negreteetal12}.

\begin{figure*}[tph]
\begin{center}
\includegraphics[scale=0.45]{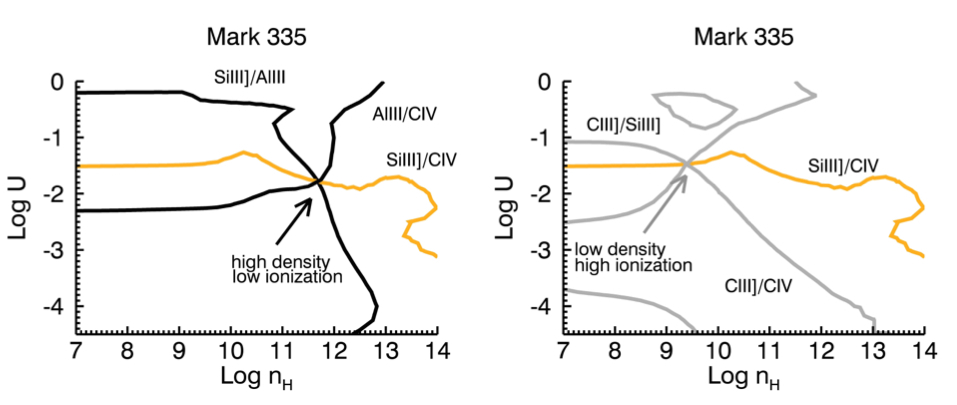}
\caption{{ The parameter plane $U$ -- \nh, with isopleths of constant emission line ratio,  for the case of Mark 335. Left panel: high density, low ionization solution; right: low density, high ionization solution.  Black and grey lines  trace ratios involving \ciii\ and \aliii\ respectively.  See text for more details. \label{fig:iso}}}  
\end{center}
\end{figure*}

At the Population B end of the 4DE1 sequence most  sources show weak optical Fe{\sc ii} emission as well as very weak or undetectable 
Al{\sc iii}$\lambda$1860 i.e.,  Al{\sc iii}$\lambda$1860 / C{\sc iii}]$\lambda$1909 $\rightarrow$ 0. Many of these sources are radio loud. 3C 390.3  shows complex line profiles including a prominent blue hump in the Balmer lines and HILs. The blue hump (which is unlikely to be the 
broad component  because of the large shift) shows  Al{\sc iii}$\lambda$1860/ C{\sc iii}]$\lambda$1909   $\approx$  0.05 while only an 
upper limit can be assigned to optical Fe {\sc ii} emission.  In this case there is no high-density solution. The measured emission line ratios 
converge toward a low-density solution that also accounts  for the observed Al{\sc iii}$\lambda$1860 / C{\sc iii}]$\lambda$1909 ratio:  
$\log (n_\mathrm{H} U) \approx 8.55 \pm 0.14$\ at 1$\sigma$. We can estimate $n_\mathrm{H}$ and  $U$ separately (not just their product): 
$\log (n_\mathrm{H}) \approx 9.95 \pm 0.13$\   and $\log (U) \approx -1.40 \pm 0.12$\ at 1$\sigma$ \citep{negreteetal14}.

\subsection{Chemical abundances of the BLR gas}
\label{metals}

The diagnostic ratio  (\siiv + \oiv) / {\sc Civ}\-$\lambda$\-1549 can be  most easily used for metallicity estimation. Other ratios like \nv / \heii\  may be stronger indicators \citep{ferlandetal96} but are more difficult to measure since e.g. \nv\ is heavily blended with \lya.  The 4DE1 sequence is likely a sequence of  ionization in the sense of a steady decrease in prominence of the low-ionization emission  towards Population B. Metallicity  might also play a  role: $Z$\ may be highly super-solar especially for  extreme Pop. A sources i.e. when \feii\ is stronger than  \hb\ then decreasing to solar or slightly super-solar values in Pop. B sources (\citealt{shinetal13,sulenticetal14}. 

\section{Radius of BLR and black hole mass}
\label{rm}

If we know {\em the product of } $n_\mathrm{H}$\ and $U$ then we  can estimate $r_{\rm BLR}$\ by inverting Eq. \ref{eq:u}:


\begin{equation}
r_{\rm BLR} = {\mathrm{const.}} {(U n_{\mathrm H})^{-\frac{1}{2}}} \left( {\int_{\nu_0}^{+\infty} \frac{L_\nu}{h\nu} d\nu} \right)^\frac{1}{2}
\label{eq:r}
\end{equation}

where the first and second bracketed factors represent the physical conditions and the number of ionizing photons respectively. This technique has 
been applied by several authors in past decades \citep{padovanietal90,wandeletal99}. The recent analysis of \citet{negreteetal13} showed 
that there is very good agreement between  $r_{\rm BLR}$\  estimates based on the photon flux and  $c\tau$ -- most likely  better than estimates 
obtained from the correlation between $c\tau$ and luminosity ($c \tau \propto L_{\lambda}^\mathrm{a}$ where $a$ = 0.5 -- 0.7, with  $a \approx 0.53$; \citealt{bentzetal13}), on which virial black hole mass estimates in high redshift quasars are based\citep[e.g.,][]{shenliu12}.

\subsection{Implications for the virial broadening estimator}

The  \hb\ profile is unshifted within $\pm $ 200 \kms\ and symmetric  in most Pop. A sources. Measurements of the \hb\ (or \mgii) line width are 
therefore considered the safest virial broadening estimator and was recently confirmed for $\approx$ 80 \% of Pop. A sources. 
Population B sources show  more complex profiles  where often shows \hb\ redshifts and redward asymmetries \citep{punsly10,marzianietal13a}. 
Shift amplitudes are however modest with shift/width ratios typically below 0.2 allowing one to extract a reasonable virial broadening 
estimator also for Pop. B sources. In order to account for non-virial motions in  the integrated line profiles it has proven effective to 
isolate three line profile components: (1) an almost unshifted and symmetric broad component  (BC) assumed to be broadened by virial  motion, (2) an occasional  blueshifted component (similar to CIV) due to gas outflows and (3) for Pop. B only, a very broad 
component (VBC) involving the \hb\ emitting region that produces no \feii. The VBC can be interpreted as arising in the 
innermost (highest ionization) part of the Balmer line emitting region \citep{huetal12}. 

A measure of the BC component width in UV  lines other than \civ\ and \lya\  is needed to facilitate black hole mass measurements for high 
redshift quasars.  The \civ\  profile width is unreliable for \mbh\ estimates, especially for Pop. A sources \citep{netzeretal07,sulenticetal07,fineetal10}, and 
\lya\ is frequently contaminated by absorption features. \citet{negreteetal13} and \citet{negreteetal14} recently compared HST archival 
spectra and optical observations for several type-1 AGNs. These authors verified that BC FWHM measures for  \hb\ are in very good agreement 
with  FWHM measures for  \siiii\ and \aliii\  that can be  considered as the best virial broadening estimators in the far UV  along with \mgii\ \citep[][and 
references therein]{trakhtenbrotnetzer12,marzianietal13a}. The virial \mbh\ mass can then be computed taking into account  that the structure factor is 
also changing along eigenvector 1 \citep{collinetal06}. 

\subsection{The case of Mark 335}

Mark 335 (PG 0003+199) is a low-$z$ quasar with archival FOS observations as well as extensive monitoring that has lead to an accurate cross 
correlation lag estimate, $\log c \tau \approx 16.61 $ [cm] \citep{kaspietal00,kollatschnyetal06,bentzetal13}.  The \civ\ and \ciii\ spectral ranges 
are shown in Fig. \ref{fig:prof}. Analysis of emission line blends is especially straightforward in this case because lines appear to be dominated by the 
virial BC with only minor  \civ\ blueshifted emission. Fig. \ref{fig:prof} shows line profiles fit with a Lorentzian function that is appropriate for 
Population A sources \citep[e.g.,][]{zamfiretal10}. Fig. \ref{fig:iso} shows the isopleths of constant diagnostic ratios. Their crossing points 
identify low density  ($\log n_\mathrm{H}U \approx 7.91 $ \ [\cm3], left) and high density solutions ($\log n_\mathrm{H}U \approx 9.94 $ \ [\cm3], right).  
The high density solution is close to the one obtained by using the measured $ c \tau $ in Eq. \ref{eq:r}, $\log n_\mathrm{H}U \approx 10.39 $ \ [\cm3]. 
Using Eq. \ref{eq:r} to estimate $\log r$\ we obtain 16.83 [\cm3] with a difference of 0.22 dex from the reverberation value \citep{negreteetal14}.

\begin{figure*}[ht]
\begin{minipage}[t]{0.75\linewidth}
\centering
\includegraphics[scale=0.35]{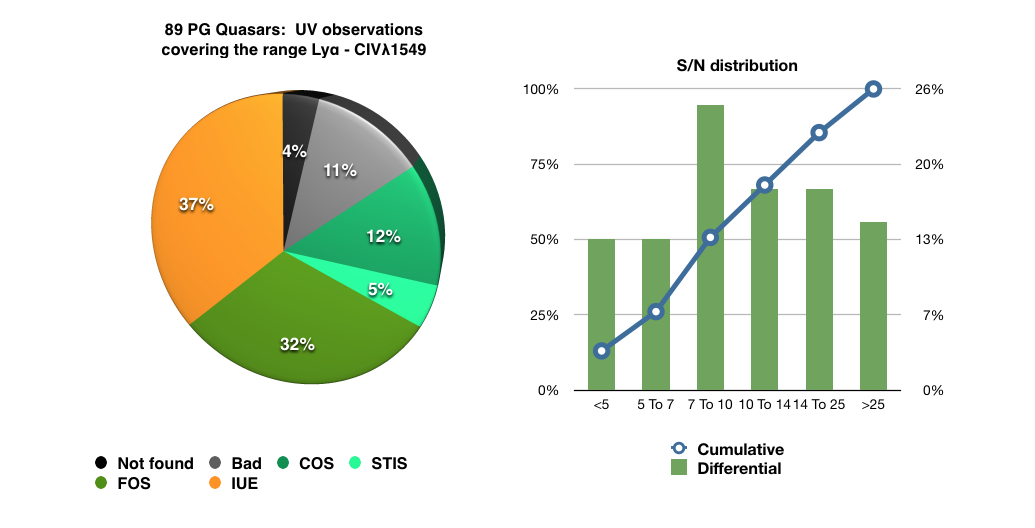}
\end{minipage}
\hspace{0.5cm}
\begin{minipage}[t]{0.2\linewidth}
\centering\vspace{-6cm}
\caption{Left: archival spectroscopic UV observations available for analysis, by instruments, for a sample of 89 PG quasars (data from \citealt{shinetal13}). Right: differential and cumulative S/N distribution for the available observations. \label{fig:mast} }
\end{minipage}
\end{figure*}

\section{MAST Inspection}
\label{mast}

{ Unbiased quasar evolution studies on  black hole mass, Eddington ratio, and chemical abundances require samples at low and high-$z$\ that match their luminosity distributions and  that cover the same rest frame range \citep{willottetal10,sulenticetal14}.}
The Mikulski Archive for Space Telescope (MAST)  provides the backbone for statistical analysis of low-$z$ AGN UV properties and for comparisons with high-$z$ 
sources. The Palomar Green \citep{greenetal86}  survey has played an important role in the analysis of optical observations for low-$z$, bright quasars. 
A search of MAST indicates that only 50\%\ of PG sources  have usable HST spectra \citep{shinetal13} while 1/6 have no data at 
all (Fig. \ref{fig:mast}). The S/N distribution also indicates that most sources have data whose quality is inferior to available optical observations.  {  Therefore, the number of low-$z$ sources with high S/N UV spectra   is insufficient to yield a pool wide enough to sample  the low-$z$\ phenomenology with high statistical significance. }  There is a  problem of completeness that is even worse if obscured/type-2 AGNs are considered \citep[e.g.,][]{roigetal14}. 
The relative scarcity of data has a strong impact on our understanding of key issues. Several 
studies still do not reach a consensus on whether metallicity $Z$ is correlated with luminosity, Eddington ratio, or black hole 
mass \citep[e.g.][and references therein]{matsuokaetal11,shinetal13}.   In addition, very few sources have been monitored in the UV 
where HIL profile changes may be associated to the high-ionization outflow properties.  
 




\
\section{Conclusion}

Rest frame UV emission line measures  are crucial to our understanding of physical conditions in quasars. 
They make possible estimates of emitting region radius, black hole mass, Eddington ratio, etc.
The 4D Eigenvector 1 formalism allows us to contextualize the diagnostic analysis of sources that are physically 
and structurally different. Archives from past /present space missions have been extremely valuable in obtaining key 
information about the inner structure of quasars, most notably strong evidence for significant outflows in a large  fraction 
of sources. Much is still needed in terms of population coverage and data quality in order to address  key 
issues related to metallicity and to outflows that ultimately may provide feedback effects relevant to the 
evolution of the  host galaxies. 

\vfill\eject
\bibliographystyle{spr-mp-nameyear-cnd}

\vfill\eject

\end{document}